\documentclass[10pt,twocolumn,letterpaper]{article}

 \usepackage{iccv}              

\newcommand{\seg}[0]{\text{seg}}
\newcommand{\ihc}[0]{\text{ihc}}
\newcommand{\gen}[0]{\text{gen}}
\newcommand{\stain}[0]{\text{stain}}
\newcommand{\he}[0]{\text{he}}
\newcommand{\bidir}[0]{\text{bidirectional}}
\newcommand{\multistain}[0]{\text{multi-stain}}

\usepackage{multirow,longtable}
\usepackage{xspace}
\newcommand{\name}[0]{{\sc StainDiffuser}\xspace}
\usepackage{rotating}

\definecolor{iccvblue}{rgb}{0.21,0.49,0.74}
\usepackage[pagebackref,breaklinks,colorlinks,allcolors=iccvblue]{hyperref}

\title{\name: MultiTask Diffusion Model for Virtual Staining}

\author{
Tushar Kataria$^{1}$ 
\and
Beatrice Knudsen$^{2}$ 
\and 
Shireen Y. Elhabian$^{1,*}$ \\
{\tt\small \{tushar.kataria@sci,beatrice.knudsen@path,shireen@sci\}.utah.edu} \\
 $^{1}$Scientific Computing and Imaging Institute \& Kahlert School of Computing \\
 $^{2}$Department of Pathology \\ University of Utah, Salt Lake City, UT, USA \\
 $^{*}${\textbf{Corresponding Author}}
}

\begin{document}
\maketitle
\begin{abstract}
Hematoxylin and Eosin (H\&E) staining is widely regarded as the standard in pathology for diagnosing diseases and tracking tumor recurrence. 
While H\&E staining shows tissue structures, it lacks the ability to reveal specific proteins that are associated with disease severity and treatment response. Immunohistochemical (IHC) stains use antibodies to highlight the expression of these proteins on their respective cell types, improving diagnostic accuracy, and assisting with drug selection for treatment.
Despite their value, IHC stains require additional time and resources, limiting their utilization in some clinical settings.
 Recent advances in deep learning have positioned Image-to-Image (I2I) translation as a computational, cost-effective alternative for IHC. I2I  generates high-fidelity stain transformations digitally, potentially replacing manual staining in IHC. 
Diffusion models, the current state-of-the-art in image generation and conditional tasks, are particularly well-suited for virtual IHC due to their ability to produce high-quality images and resilience to mode collapse.
However, these models require extensive and diverse datasets (often millions of samples) to achieve a robust performance, a challenge in virtual staining applications where only thousands of samples are typically available.
Inspired by the success of multitask deep learning models in scenarios with limited data, we introduce \name, a novel multitask diffusion architecture tailored to virtual staining that achieves convergence with smaller datasets. 
\name simultaneously trains two diffusion processes: (a) generating cell-specific IHC stains from H\&E images and (b) performing H\&E-based cell segmentation, utilizing coarse segmentation labels exclusively during training. Our results demonstrate that \name produces high-quality virtual staining results for two markers, CK8/18 (epithelial cell marker) and CD3 (T-lymphocyte marker), out-performing more than twenty I2I baselines. 
\href{To be Released on Acceptance}{ All codes and associated trained models will be publicly released via github upon acceptance.}
\end{abstract}
\section{Introduction}
Hematoxylin and Eosin (H\&E) staining is a key tissue staining technique in pathology. H\&E reveals features of tissue architecture, cellular organization, and nuclear morphology that allow pathologists to diagnose diseases and guide treatment decisions \cite{pai2022measuring,park2016histological}. However, H\&E staining does not reveal cell differentiation and activation states that are critical for cancer subtyping, assessment of cancer severity and treatment with targeted drugs \cite{liu2022bci,magaki2019introduction,knudsen2009novel}. 
To gain information from tissues that cannot be discerned in H\&E stained tissues, pathologists use immunohistochemical (IHC) staining methods. IHC uses antibodies to mark specific proteins that highlight cell types or functional cell states, which cannot be identified using H\&E alone. For example, HER2 and Ki67 are used for breast cancer subtyping \cite{li2023adaptive,liu2022bci}, CDX2/CK818 to help confirm colon cancer metastases \cite{baba2009relationship}, and CD3 or CD20 localize T-cells or B-cells, respectively \cite{meddens2018biophysical}. 
IHC is widely used in clinical settings to provide information beyond the tissue architecture and nuclear morphology captured by H\&E stains \cite{magaki2019introduction}.
 However, despite the critical need of IHC stains, the staining process is labor-intensive, costly, and time-consuming \cite{magaki2019introduction,rahman2022alcoholic,latonen2024virtual}.

Virtual staining \cite{liu2022bci,dubey2024vims,dubey2023structural,li2023adaptive,bian2024hemit}, powered by deep learning, provides a rapid and cost-effective alternative to manual, laboratory-based IHC staining. In addition, virtual staining allows standardization, reduces stain variability, and enhances consistency in pathology image analysis \cite{latonen2024virtual}. Prior to their application for virtual IHC, deep learning models have been applied to easier tasks, such as H\&E staining of unstained tissues, H\&E to chemical stain conversion in renal pathology \cite{ozyoruk2021deep, ho2024f2fldm}, fluorescence-to-H\&E conversion \cite{wang2019deep, rivenson2019virtual}, H\&E-to-mIF conversion \cite{bian2024hemit}, and other related tasks \cite{ghahremani2022deepliif}. Unlike these simpler image translation tasks that leverage morphological cues, our work addresses the more challenging conversion from brightfield H\&E to brightfield IHC. We propose virtual staining models to highlight CD3-positive T-cells —lymphocyte subclasses indistinguishable in H\&E-stained tissue sections. Since T-cells cannot be differentiated by morphology alone, IHC is essential for accurate identification. Training a virtual staining model for this task is particularly challenging, requiring the learning of complex, non-linear features of both the cells and their microenvironment.

Diffusion models have emerged as the state-of-the-art in a wide range of generative and conditional-generative tasks, including text-to-image generation \cite{zhang2023text,baldridge2024imagen,brooks2023instructpix2pix}, inpainting \cite{saharia2022palette}, colorization, and super-resolution \cite{saharia2022image}. These models have demonstrated exceptional performance in learning the underlying data distributions, particularly when provided with ample training data and resources \cite{blattmann2023stable,wang2025vidprom}. Diffusion models consistently outperform GAN-based models in the diversity and quality of generated images, making them a powerful tool in generative tasks.
However, diffusion models require large training datasets  (millions of samples) to converge and effectively learn the underlying data distribution \cite{cao2024survey,hur2024expanding}. Their performance is subpar in virtual staining tasks, especially when training data is scarce, and dataset samples are typically in the thousands rather than millions \cite{hur2024expanding,abraham2023comparison}. 
Because of this deficiency, we propose a Multitask Diffusion model framework for our virtual staining application. Multitask deep neural networks have been shown to consistently outperform single-task models, particularly in data-limited settings and when tasks have strong interrelations \cite{ishibashi2022multi, bollmann2018multi}. Building on this insight, we introduce \name, a multitask diffusion model designed to simultaneously perform cell segmentation (\textit{task 1}) and virtual staining (\textit{task 2}). By focusing on the same cells, \name harnesses the intrinsic task affinity that significantly improves the  virtual staining quality. The collaborative relationship between segmentation and virtual staining allows \name to learn distinct features beyond mere color replication, thereby enhancing the fidelity of virtual staining. 
For training, \name utilizes H\&E and IHC images tiles that are registered at pixel level accuracy. In the IHC stained image, the DAB (diaminobenzidine) channel produces a brown signal that highlights the cell type of interest. Segmentation masks are automatically generated by thresholding the DAB channel \cite{kataria2023automating}, highlighting the target cells and eliminating the need for manual annotations.
For inference, \name relies solely on the conditioned virtual staining diffusion process based on H\&E, without requiring the IHC segmentation input. 
The proposed architecture is also robust and adaptable, enabling the simultaneous generation of multiple IHC stains from the same H\&E (i.e., multiplexed staining).

Although I2I translation models show promise for virtual staining, no comprehensive benchmark exists for current state-of-the-art I2I models. Most studies \cite{li2023adaptive,liu2022bci,dubey2023structural} focus on a few models, such as Pix2Pix \cite{isola2017image}, CycleGAN \cite{zhu2017unpaired}, or CUT \cite{park2020contrastive}, providing limited insights into their strengths and weaknesses.
We address this literature gap by presenting the first large-scale comparison of over twenty I2I models, including GAN-based and diffusion-based architectures. Our results establish a new benchmark for evaluating models in IHC virtual staining tasks (see Table \ref{tab:results_main_table}), facilitating informed model selection and development in the future.
\noindent The main contributions of this manuscript are:
\begin{itemize}
   \item \name, a novel multitask diffusion architecture for I2I translation demonstrates its effectiveness in the challenging virtual IHC staining of T-cells, that cannot be easily identified by morphological cues alone.
   \item This architecture can also simultaneously generate multiple stain types (i.e., multiplexed staining), providing enhanced versatility for various staining tasks. 
    \item We present the first comprehensive evaluation of over twenty GAN-based and diffusion-based I2I models for virtual staining, establishing a new benchmark.
    \item We provide a comprehensive analysis using qualitative and quantitative metrics (SSIM, PSNR, FID, KID, Precision, Recall) on two new datasets.
\end{itemize}
\section{Related Works}

\textbf{Conditional Diffusion Models.}
Diffusion models \cite{ho2020denoising} represent the current state of the art in many unconditional generation \cite{dhariwal2021diffusion} and conditional generation tasks such as inpainting, colorization, superresolution and compression \cite{saharia2022palette,yang2023diffusion,saharia2022image,fuest2024diffusion}. 
In medical applications, conditional diffusion models,  have been effectively applied to segmentation \cite{kazerouni2022diffusion,pinaya2022fast,wolleb2022diffusion,wu2024medsegdiff}, reconstruction \cite{chung2022mr,wang2023inversesr}, and registration tasks \cite{kim2022diffusemorph}. More recently, diffusion and latent diffusion models have gained traction in medical image generation, with applications in volumetric data augmentation \cite{guo2024maisi}, H\&E synthesis \cite{ho2024f2fldm}, and text-conditioned virtual staining \cite{dubey2024vims}. Despite advances, diffusion models have seldom been used for H\&E-conditioned IHC virtual staining due to their large data requirements \cite{wang2024patch} and challenges with small datasets, limiting their use to data-rich areas \cite{abraham2023comparison,hur2024expanding}. In this work, we propose a multitask diffusion architecture tailored to improve performance on smaller datasets for H\&E-conditioned virtual staining.

\textbf{Multi-Task Deep Neural Networks}. Multitask deep neural networks have demonstrated superior performance compared to single-task models when tasks share a high degree of affinity \cite{fifty2021efficiently, jiang2024forkmerge, crawshaw2020multi, yu2020gradient}. Notably, these multitask models also outperform single-task counterparts in data-constrained scenarios \cite{ishibashi2022multi, bollmann2018multi, karanam2023adassm}. Diffusion models trained across multiple tasks have also shown strong performance in areas like depth prediction \cite{zhou2020pattern}, tumor growth forecasting \cite{wolleb2022swiss}, and other vision tasks \cite{ye2024diffusionmtl}. Our approach shares the intuition of using multitasking to regularize the loss landscape, especially beneficial with limited data (thousands rather than millions of samples). Closely related to our work is ControlNet \cite{zhang2023adding}, which introduces model parameters to spatially condition the generation process of a Latent Diffusion Model (LDM). In contrast, we employ two separate diffusion processes that interact through a shared encoder to achieve similar regularization. This separation allows each diffusion process to focus on its specific task while benefiting from cross-task interactions via attention maps, enhancing the overall learning efficiency. Our approach also bears similarity to DiffusionMTL \cite{ye2024diffusionmtl}, which uses separate diffusion models and multitask conditioning to refine the predictions of a multitask backbone network. However, unlike DiffusionMTL, our architecture directly generates outputs using multiple diffusion processes without relying on a multitask model for prediction and diffusion for refinement. This design minimizes complexity and enables early task interaction in latent space, which has been shown to outperform the late interaction approach used in DiffusionMTL \cite{nishi2024joint,gadzicki2020early,vandenhende2020mti,li2022learning}. While our method shares some conceptual similarities with DeepLIIF \cite{ghahremani2022deepliif} and DiffI2I\cite{xia2024diffi2i} in using segmentation and generation as auxiliary task, which validates the affinity between these tasks, there are several key differences. We focus on H\&E to Brightfield IHC conversion, the inverse of DeepLIIF \cite{ghahremani2022deepliif} task, making our approach more generalizable since H\&E is widely available, whereas IHC is less accessible. Unlike DeepLIIF \cite{ghahremani2022deepliif} which performs sequential generation and segmentation without shared encoders or attention—where information sharing occurs only through loss calculation—our architecture processes these tasks in parallel. By leveraging shared encoders and latent space interaction via attention mechanisms, our method establishes a more powerful and integrated paradigm.  DiffI2I \cite{xia2024diffi2i} is more closely to LDM\cite{rombach2022high,ho2024f2fldm} and BBDM/LBBDM\cite{li2023bbdm} where it uses single diffusion process with two losses, whereas we use separate diffusion processes for segmentation and generation, with attention parameters indirectly influencing each other via the shared (H\&E) encoder. Additionally, our segmentation masks are generated automatically through thresholding (refer to section 4) \cite{kataria2023automating}, whereas other methods rely on manual segmentations, making our approach more efficient.

\textbf{Image-to-Image (I2I) Translation models.} 
I2I models can be classified into two categories: (a) \textit{paired models}, where images from two domains are pixel-aligned, enabling pixel-level supervision, such as two stains on the same tissue (requiring registration for alignment), and (b) \textit{unpaired models}, where images from the domains lack direct alignment, such as stains on tissues from different patients or different tissues from the same patient (with no pixel alignment possible, even with registration).
Pix2Pix \cite{isola2017image} and CycleGAN \cite{zhu2017unpaired} are the two notable works for conditional image generations for \textit{paired} and \textit{unpaired} datasets, respectively. 
Other methods build upon these models with domain-specific modifications, such as multi-resolution techniques in PyramidPix2Pix \cite{liu2022bci} and various regularization strategies, including CUT/FastCUT \cite{park2020contrastive}, AdaptiveSupPatchNCE \cite{li2023adaptive}, SANTA \cite{xie2023unpaired}, UNSB \cite{kim2023unpaired}, StegoGAN for reducing hallucinations \cite{wu2024stegogan}, SC-GAN, which adds edge generation as a regularization \cite{dubey2023structural}, and UVCGAN, which incorporates a pre-trained encoder \cite{torbunov2023uvcgan}. Numerous other methods explore similar enhancements \cite{liu2017unsupervised,kim2019u,zhao2020unpaired,Chen_2020_CVPR,tang2021attentiongan,hu2022qs,xieunsupervised, chen2022eccv}.
The performance of paired architectures depends on the size of the dataset, as the number of paired samples is limited to \( O(N) \), where \( N \) is the number of images within the domain. In contrast, models designed for unpaired settings have access to a much larger number of training samples, scaling as \( O(N^2) \). However, GAN-based models can suffer from hallucinations if the images from the two domains are not well-aligned \cite{isola2017image, liu2022bci,kong2021breaking,honkamaa2023deformation,klages2020patch}, or may conceal domain-specific information in high-frequency texture features, leading to hallucinations \cite{wu2024stegogan}. This raises concerns about their reliability, particularly for medical applications. 
GAN-based models are susceptible to mode collapse and reduced sample diversity, while diffusion models offer improved distribution coverage and higher sample quality in comparison \cite{xiao2021tackling,durall2020combating,thanh2020catastrophic,zhao2018bias,yang2019diversity}. 

Our approach builds on diffusion models designed for paired domain translation. We propose a novel and flexible architecture that integrates diffusion models with multitasking for conditional generation, making it highly effective for applications with limited dataset sizes. Although we demonstrate its effectiveness in virtual staining, the architecture is versatile and can be applied to a wide range of vision tasks, such as segmentation, depth prediction, and MRI-to-CT \cite{ye2024diffusionmtl,guo2024maisi}. 
We leverage our architecture's flexibility to enable virtual multiplexing, generating multiple stains from a single H\&E image with uniplex supervision—pairing H\&E patches with a single IHC stain. This capability distinguishes our method, as no existing techniques offer it.

\section{Diffusion Model Background}
The denoising diffusion probabilistic model (DDPM) defines the diffusion process as a two-stage process, i.e., forward and reverse. 
The \textit{forward process} progressively corrupts the image, $\textbf{I}_0 \in \mathbb{R}^{H \times W}$, by adding Gaussian noise over $T$ iterations in a Markovian manner. 
A neural network parameterized by $\mathbf{\theta}$, denoted as $f_{\theta}( \textbf{I}_t(\mathbf{I}_0,\Bar{\alpha}_t),t)$, takes the noisy image $\mathbf{I}_t$ and the current random noise level $\Bar{\alpha}_t$ as inputs to estimate the noise vector $\boldsymbol{\epsilon}$ used to corrupt the original image $\mathbf{I}_0$. The loss function is defined as the mean squared error (MSE) between the estimated noise and the original noise,
\vspace{-0.5em}
\begin{align}
    \mathcal{L}(\theta) &= \mathbb{E}_{t, \mathbf{I}_0, \boldsymbol{\epsilon}} \left[||\boldsymbol{\epsilon}-f_{\theta}( \textbf{I}_t(\mathbf{I}_0,\Bar{\alpha}_t),t)||^2 \right] 
\end{align}

In conditional diffusion models, the network, parameterized by $\theta$, is also conditioned on the input image $\mathbf{I}_x$ or a latent space representation from a pre-trained model, resulting in an estimation function defined as
$f_{\theta}( \mathbf{I}_x, \textbf{I}_t(\mathbf{I}_0,\Bar{\alpha}_t),t)$. The loss function for conditional diffusion models is given by:
\vspace{-0.5em}
\begin{align}
    \mathcal{L}_{\text{cond}}(\theta) &= \mathbb{E}_{t, (\mathbf{I}_x, \mathbf{I}_0), \boldsymbol{\epsilon}} \left[||\boldsymbol{\epsilon}-f_{\theta}( \mathbf{I}_x, \textbf{I}_t(\mathbf{I}_0,\Bar{\alpha}_t),t)||^2 \right] 
\end{align}
For a detailed background on diffusion models, please refer to the supplementary materials section \ref{apdx:diffusion_model_background}.

\begin{figure*}[!thb]
    \centering
    \includegraphics[width=0.95\linewidth]{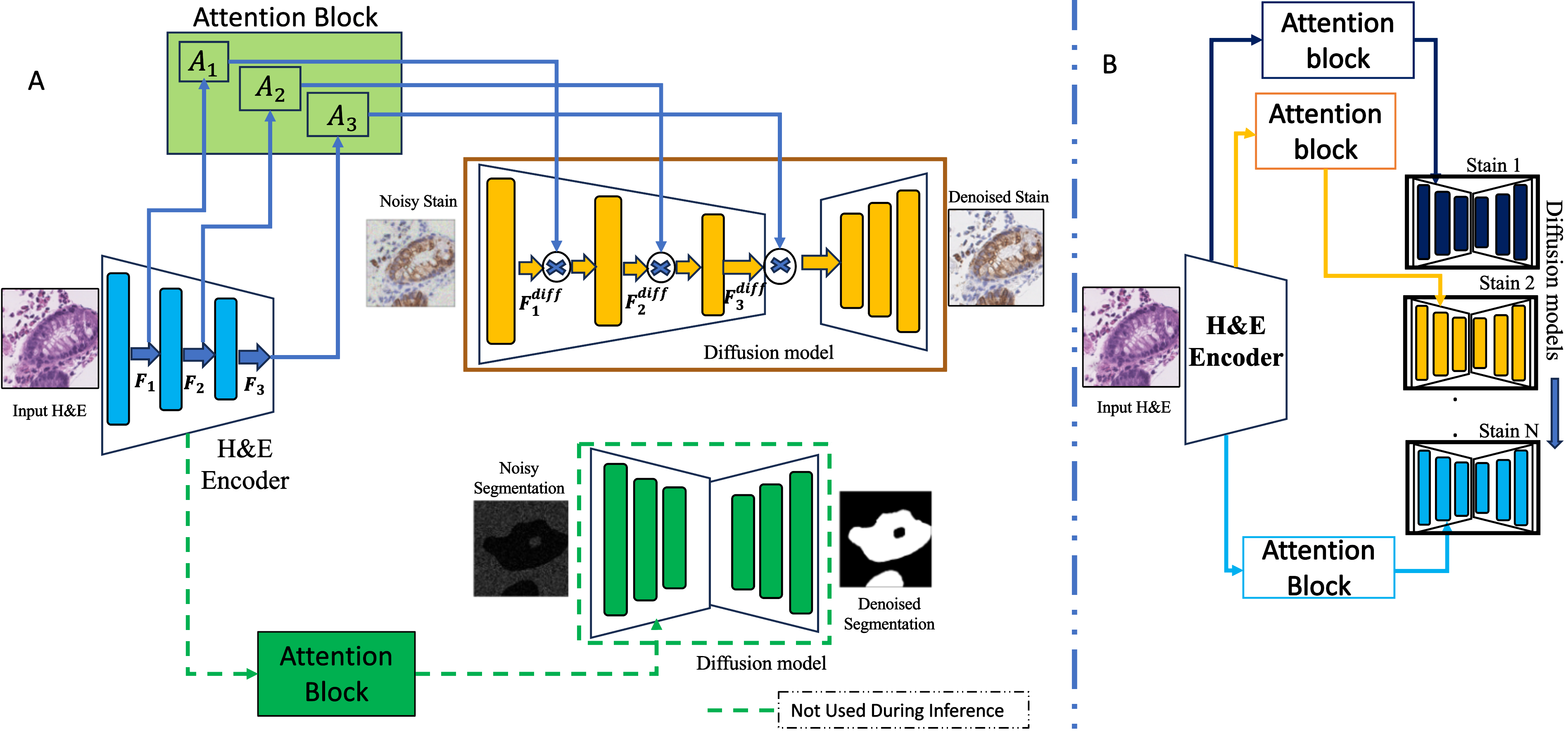}
    \caption{\textbf{\name}: (A) Block diagram of the \name model. The two diffusion models are tasked to (a) denoise the noisy IHC and, (b) corresponding segmentation of the cells on H\&E images. The two diffusion models (with separate parameters) interact through the common H\&E encoder and the attention blocks for different tasks as well as the back-propagation of losses.  (B) Block Diagram of \name extension for Multi-staining task, with $N$ number of stains for generation.
    }
    \label{fig:Architecture}
    \vspace{-1.5em}
\end{figure*}

\section{Methods: \name }
We introduce \name, a novel architecture designed for simultaneous segmentation and virtual staining, with an extension for multi-stain generation using single-stain paired datasets. Furthermore, we adapt \name for scenarios lacking segmentation data by proposing a bi-directional multi-tasking diffusion variant.

 \textbf{\textbf{\name Architecture}:}
The architecture of the proposed model is shown in Figure \ref{fig:Architecture}A. 
The proposed multitask framework consists of two tasks: (a) virtual staining of H\&E images (generation diffusion model) and (b) H\&E cell/object segmentation (segmentation diffusion model). The segmentation task learns to identify the same cells/objects that the virtual staining model is learning to color, establishing an implicit affinity between the tasks. The two diffusion branches share information indirectly via the input image encoder (H\&E), with an attention block that facilitates mutual learning during the training process. 
The proposed architecture consists of two diffusion processes trained simultaneously. It includes an H\&E encoder and two UNet\footnote{The UNet architecture is commonly used for diffusion models.} diffusion networks, parameterized by $\theta$ and $\phi$, for segmentation and generation, respectively.

The H\&E encoder processes the input image, generating feature maps at different levels of abstractions 
 $F_1, F_2, F_3$ and $F_4$ (Figure \ref{fig:Architecture}A only shows three features for simplicity). These features are used to compute attention matrices $(A^{Task}_1,A^{Task}_2,A^{Task}_3,A^{Task}_4)_{Task \in {seg,gen}}$. Separate attention matrices are computed for each task (generation and segmentation), with each attention block using distinct parameters (no parameter sharing). This ensures that the tasks remain specialized while interconnected through the shared H\&E encoding space, creating a stronger interaction between the generation and segmentation branches beyond the loss functions. Similarly, the diffusion model's encoder generates features of matching dimensions($F_1^{diff} -F_4^{diff}$). During the forward pass, the attention matrices $(A_1-A_4)^{Task}_{{seg,gen}}$ are used to modulate corresponding diffusion features as shown in Figure \ref{fig:Architecture}A, effectively integrating the H\&E features in diffusion process. The decoder of the diffusion block is a simple unit-decoder with skip connections.

Let $\textbf{I}^\he$ represent the input H\&E image, $\textbf{I}^{\seg}_t$ the noisy segmentation image at time step $t$, and $\textbf{I}^{\ihc}_t$ the noisy IHC image given to the segmentation and generation branches, respectively. The segmentation diffusion branch, parameterized by $\theta$, is trained to minimize the following loss:
\vspace{-0.5em}
\begin{align*}
    \mathcal{L}_{\seg}(\theta) &= \mathbb{E}_{t, (\mathbf{I}^\he,\mathbf{I}_0^{\seg}), \boldsymbol{\epsilon}} \left[||\boldsymbol{\epsilon}-f_{\theta}( \mathbf{I}^\he, \textbf{I}^{\seg}_t(\mathbf{I}_0^{\seg},\Bar{\alpha}_t),t)||^2 \right] 
\end{align*}

Similarly, the generation diffusion branch, parameterized by $\phi$, is trained to minimize:
\vspace{-0.5em}
\begin{align*}
    \mathcal{L}_{\gen}(\phi) &= \mathbb{E}_{t, (\mathbf{I}^\he,\mathbf{I}_0^{\ihc}), \boldsymbol{\epsilon}} \left[||\boldsymbol{\epsilon}-f_{\phi}( \mathbf{I}^\he, \textbf{I}^{\ihc}_t(\mathbf{I}_0^{\ihc},\Bar{\alpha}_t),t)||^2 \right] 
\end{align*}

\name is trained to minimize the sum of these losses, specifically,
 $   \mathcal{L}_{\stain} = \mathcal{L}_{\gen} + \lambda \mathcal{L}_{\seg}$
, where $\lambda$ is a hyperparameter that controls the interaction between tasks, regulating their impact on one another.

\textbf{Segmentation Annotations:} Manual segmentation of cells or objects significantly increases deployment costs and limits model scalability due to the expertise required by pathologists for accurate annotation. To reduce the need for manual annotations, we use coarse segmentations generated through thresholding and morphological operations, similar to \cite{kataria2023automating}. This automated approach eliminates the need for manual labeling, expanding the applicability of the proposed architecture. Importantly, these coarse segmentations are only used during training, allowing the generation diffusion model to operate independently during inference without any need for segmentation inputs, see Figure \ref{fig:Architecture}.

\textbf{Multi-Stain/Multiplex \name:} The proposed \name architecture is inherently flexible, supporting seamless adaptation for diverse tasks. This adaptability enables it to generate multiple stains from a single H\&E image. Specifically, as the interaction between the H\&E encoder and the diffusion models is facilitated by distinct attention networks, allowing the same H\&E encoder to produce multiple stains concurrently. Figure \ref{fig:Architecture}B illustrates the proposed multi-stain generation framework. For training this architecture, we leverage paired, uniplex dataset containing only pairs of $(\mathbf{I}^{\he_i}, \mathbf{I}^{\ihc_i})$ for each IHC stain, rather than multiplexed data, which includes all stains for each H\&E image $(\mathbf{I}^{\he}, \mathbf{I}^{\ihc_1}, \mathbf{I}^{\ihc_2}, \cdots, \mathbf{I}^{\ihc_N})$. If $N$ represents the number of stains, the training loss is defined as follows:
\vspace{-0.5em}
\begin{align*}
    \mathcal{L}_{\multistain}(\phi_1, \cdots, \phi_N) = & \nonumber\\
    \sum_{i=1}^N \lambda_i \mathbb{E}_{t, (\mathbf{I}^{\he_i},\mathbf{I}_0^{\ihc_i}), \boldsymbol{\epsilon}} &\left[||\boldsymbol{\epsilon}-f_{\phi_i}( \mathbf{I}^{\he_i}, \textbf{I}^{\ihc_i}_t(\mathbf{I}_0^{\ihc_i},\Bar{\alpha}_t),t)||^2 \right] 
\end{align*}

The attention blocks enable diffusion models to focus on stain-specific features in IHC images, capturing complementary information that enhances learning across stains.

\textbf{\textbf{Bi-Directional Multi-task Diffusion}:} 
The architecture described above relies on coarse segmentation, which may not be available for all stains. To enhance usability, we propose a multi-task variant that eliminates the need for segmentation during training while still benefiting from multi-task learning. In this variant, the segmentation diffusion model is removed from \textit{\name}, and the two tasks are trained using a single-generation diffusion model. The training tasks are: (a) H\&E to IHC generation and (b) IHC to H\&E generation. This model features a single input image encoder (alternating between H\&E and IHC inputs) and a single U-Net diffusion model, parameterized by $\theta$, to handle both diffusion processes. In the first diffusion process, H\&E is provided as input to the encoder, and the model is tasked with denoising the noisy IHC. In the second diffusion process, the roles of H\&E and IHC are reversed.
Let input H\&E and IHC images be denoted as $\textbf{I}^{\he}$ and $\textbf{I}^{\ihc}$, respectively, and noisy H\&E and IHC at time step $t$ as $\textbf{I}^{\he}_t$ and $\textbf{I}^{\ihc}_t$, respectively. The final loss used to train the bidirectional model is:
\vspace{-0.5em}
\begin{align*}
    \mathcal{L}_{\bidir}(\phi) =  \mathbb{E}_{t, (\mathbf{I}^\he,\mathbf{I}_0^{\ihc}), \boldsymbol{\epsilon}} \left[||\boldsymbol{\epsilon}-f_{\phi}( \mathbf{I}^\he, \textbf{I}^{\ihc}_t(\mathbf{I}_0^{\ihc},\Bar{\alpha}_t),t)||^2 \right] \nonumber\\
    + \mathbb{E}_{t, (\mathbf{I}^\ihc,\mathbf{I}_0^{\he}), \boldsymbol{\epsilon}} \left[||\boldsymbol{\epsilon}-f_{\phi}( \mathbf{I}^\ihc, \textbf{I}^{\he}_t(\mathbf{I}_0^{\he},\Bar{\alpha}_t),t)||^2 \right] 
\end{align*}
The bi-directional task design fosters a domain-invariant latent space, improving generative quality and robustness.
\section{Experimental Details and Discussion}
\textbf{Datasets.} 
We evaluated the proposed approach using two virtual staining datasets: H\&E to CD3 and H\&E to CK8/18. Each dataset consists of 92 H\&E WSIs paired with corresponding IHC stains (CK8/18 or CD3) from the same tissue samples collected during surveillance colonoscopies of patients with ulcerative colitis. 
Training patches were extracted from 70 WSIs, reserving the rest for testing. Each 256x256 patch was randomly sampled, including only those with at least 50\% tissue content. For test WSIs, non-overlapping patches were sampled sequentially. The CK8/18 dataset included 57,887 training and 6,460 test patches, while the CD3 dataset had 59,362 training and 6,920 test patches. \footnote{\small The de-identified dataset will be publicly available subject to signing a Data Transfer Agreement with our institution.}

 \textbf{Baselines Comparisons.} 
A comprehensive qualitative and quantitative comparison is performed across 20+ methods, including eight paired and 15 unpaired architectures, with five paired methods using diffusion-based models and LDM variants LBBDM-f4 \cite{li2023bbdm} (see Table \ref{tab:results_main_table}).
 Baselines, \name, and its variants (Bidirectional-\name and Multi-\name) were trained for 3 million iterations with a batch size of 4, or 200 equivalent epochs. Wherever possible, 20\% of training data was used for validation. \name models were trained on downsampled images of 64x64 and 128x128. For CD3, we used 2,000 diffusion steps, and for CK8/18, 4000 steps, were selected through hyperparameter tuning. 
 
 \textbf{Ablation Experiments.} We conduct through ablations to study the impact of various choices made for \name, (1) the number of diffusion steps during training, (2) the $\lambda$ hyperparameter, (3) number of training data samples, and (4) auxillary task type(segmentation vs Signed Distance prediction). Previous histopathology studies \cite{graham2019hover, naylor2018segmentation} have shown that predicting signed distance functions instead of segmentation masks enhances performance in cell-level tasks. Therefore, this ablation is particularly relevant to our proposed \name. 
\begin{table*}[!htb]
\scalebox{0.85}{
\setlength{\tabcolsep}{4.8pt}
    \centering
    \begin{tabular}{p{0.3cm}|c||c|c||c|c|c|c||c|c||c|c|c|c}
    & & \multicolumn{6}{c||}{\bf CK818} &  \multicolumn{6}{c}{\bf CD3}
    \\
    \hline
   & & \multicolumn{2}{c||}{\bf Texture Metrics } &  \multicolumn{4}{c||}{\bf Distribution Metrics} 
   & \multicolumn{2}{c||}{\bf Texture Metrics } &  \multicolumn{4}{c}{\bf Distribution Metrics}  \\
    \hline
   & I2I Methods & PSNR $\uparrow$ & SSIM $\uparrow$ & FID $\downarrow$& KID $\downarrow$ & Prec $\uparrow$ & Rec $\uparrow$  & PSNR $\uparrow$ & SSIM $\uparrow$ & FID $\downarrow$& KID $\downarrow$ & Prec $\uparrow$ & Rec $\uparrow$ \\
    \hline
     \multirow{16}{=}{\begin{sideways}\textbf{Unpaired}\end{sideways}}   
   &    CycleGAN  \cite{zhu2017unpaired}           & 
       \bf 20.4 & \textcolor{red}{0.66}  & \bf 23.85 &\bf 0.002 & \textcolor{red}{0.886} & \bf 0.866 &
      \textcolor{red}{19.78} & \bf 0.622  & 18.19 & 0.0029 & \bf 0.854 & 0.858 \\
   &    UNIT       \cite{liu2017unsupervised}                       & 
       12.61 & 0.036 & 28.22 & 0.0056 & 0.839 & 0.8 &
       17.63 & 0.586 & 28.84 & 0.0043 & 0.766 & 0.700\\
   &    UGATIT   \cite{kim2019u}                                    & 
       \textcolor{red}{20.24} & \bf 0.665  & 30.81 & 0.0065 & 0.787 & 0.805 &
       19.49 & 0.618 & 18.82 & 0.003 & 0.838 & 0.816 \\
   &    CUT       \cite{park2020contrastive}       & 
       19.84 & 0.645 & \textcolor{red}{25.32} & 0.0032 & 0.877 & 0.837 & 
       \bf 19.86 & 0.6137 & \bf 16.98 & \bf 0.0015 & 0.847 & \bf 0.887 \\
   &    FastCUT    \cite{park2020contrastive}                       & 
       19.35  & 0.637 & 29.6 & 0.0078 & 0.805 & 0.788 &
       \bf 19.86 & 0.615 & 25.86 & 0.0098& 0.761 & 0.86\\
   &    ACL GAN    \cite{zhao2020unpaired}                          & 
       15.06 & 0.599   & 132.5 & 0.114 & 0.042 & 0.05 &
       16.39 & 0.539 & 30.22 & 0.0139 & 0.823 & 0.671\\
   &    NICE GAN  \cite{Chen_2020_CVPR}                             & 
       19.54 & 0.65  & 26.71 & 0.0036 & 0.873 & 0.825 & 
       19.24 & 0.614 & 24.26 & 0.0075 & 0.759 & 0.804 \\
   &    Attn. GAN \cite{tang2021attentiongan}                   &  
       19.85 &  0.657 & 26.71 & 0.0036 & 0.873 & 0.825 &
       19.72 & 0.622 & 18.54 & 0.0027 & 0.849 & 0.847 \\
   &    QS GAN \cite{hu2022qs}                         & 
       19.85 &  0.643 & 25.40 & \textcolor{red}{0.0029} & 0.866 & \textcolor{red}{0.840} &
       19.75 & 0.612 & \textcolor{red}{17.34} & \textcolor{red}{0.0017} & 0.847 & \textcolor{red}{0.87}\\
   &    Decent        \cite{xieunsupervised}                        & 
       20.11  & 0.634 & 26.19 & 0.003 & 0.863 & 0.835 &
       19.54 & 0.596 & 17.64 & 0.0021 & 0.833 & 0.861 \\
   &    VQ-I2I-Un       \cite{chen2022eccv}                   & 
       12.71  & 0.087 & 41.37 & 0.0127 & 0.527 & 0.661 &
       18.53 & 0.398 & 34.30 & 0.0139 & 0.532 & 0.722\\
   &    UVCGAN    \cite{torbunov2023uvcgan}                         & 
       19.74 & 0.65 & 36.83 & 0.0124 & 0.726 & 0.771 & 
  19.85 & 0.62 & 21.98 & 0.0051 & 0.814 & 0.751 \\
   &    SANTA     \cite{xie2023unpaired}                            & 
   19.42 & 0.605 & 27.21 & 0.0039 & 0.843 & 0.786 &
   19.23 & 0.588 & 18.43 & 0.0031 & 0.838 & 0.838 \\
   &    UNSB      \cite{kim2023unpaired}                            & 
       14.34 & 0.147 & 28.72 & 0.0059 & 0.837 & 0.617 &
       15.51 & 0.17 & 27.80 & 0.013 & 0.8137 & 0.673\\
   &    StegoGAN \cite{wu2024stegogan}                              & 
       19.9 & 0.659 & 26.17 & 0.0031 & \bf 0.887 & 0.820 &
       19.74 & \textcolor{red}{0.621} & 18.49 & 0.0029 & \textcolor{red}{0.852} & 0.845 \\       
       \hline
       \hline
    \multirow{9}{=}{\begin{sideways}\textbf{Paired}\end{sideways}}   
   &    Pix2Pix    \cite{isola2017image}           & 
       19.22 & 0.555    & 57.04 & 0.0223 & 0.757 & 0.744 &
       19.38 & 0.54 & 29.00 & 0.01 & 0.656 & 0.801 \\
   &    PyramidPix2Pix \cite{liu2022bci}           & 
       20.43 &  \textcolor{red}{0.650} & 44.82 & 0.0206 & 0.822 & 0.707 &
       \textcolor{red}{20.48} & \textcolor{red}{0.612} & 40.46 & 0.026 & 0.777 & 0.661 \\
   &    VQ-I2I-P       \cite{chen2022eccv}                     & 
       12.68  & 0.081 & 45.38 & 0.0195 & 0.503 & 0.590 &
       16.67 & 0.33 & 116.4 & 0.07 & 0.121 & 0.515 \\
   &    \textcolor{blue}{EDSDE}       \cite{zhao2022egsde}                            & 
   15.02 & 0.311 & 112.4 & 0.089 & 0.144 & 0.554 &
   14.69 & 0.29 & 69.80 & 0.0564 & 0.194 & 0.194 \\
  &    \textcolor{blue}{CycleDiffusion} \cite{wu2022unifying}                        & 
  15.93 & 0.520 & 72.28 & 0.056 & 0.761 & 0.398 &
  15.47 & 0.48 & 65.02 & 0.0 545 & 0.342 & 0.754 \\
   &    \textcolor{blue}{BBDM}    \cite{li2023bbdm}                                   & 
       \textcolor{red}{20.47} & \textcolor{red}{0.65} & 33.31 & 0.0106 & 0.828 & 0.815 &
       20.14 & 0.617 & 30.31 & 0.0147 & 0.735 & 0.795 \\
   &    \textcolor{blue}{LBBDM-F4}    \cite{li2023bbdm}                         & 
       18.8 & 0.526 & 32.35 & 0.0096 & 0.731 & 0.786 &
       18.73 & 0.495 & 25.53 & 0.0084 & 0.558 & 0.806 \\
   &    \textcolor{blue}{LBBDM-F16}    \cite{li2023bbdm}                        & 
       16.46 & 0.303 & 68.67 & 0.0444 & 0.634 & 0.288 &
       17.16 & 0.311 & 70.75 & 0.0524 & 0.558 & 0.190 \\
     &   \name (Ours)                                    &  
     \bf 21.67 &	\bf 0.676 & \bf 23.46 &	\bf 0.0028 &	\bf 0.933 & \bf 0.926 &
   \bf   20.64    &  \bf 0.633 & \bf 15.83 & \bf 0.0012 & \bf 0.863 & \bf 0.905 \\
   &    \textit{BiDirec}-{\sc StainD}   (Ours)                  &  
   \textcolor{red}{20.47}	& \textcolor{red}{0.650}	& \textcolor{red}{25.02} &	\textcolor{red}{0.0047}	& \textcolor{red}{0.914}	& \textcolor{red}{0.880} &
   20.08 & 0.600 & 19.22 & 0.0034 & 0.779 & \textcolor{red}{0.891} \\
   \hline
   \multirow{2}{=}{\begin{sideways}\textbf{Quasi}\end{sideways}}  
   &    \textit{Multi}-{\sc StainD}  (Ours)                         &  
  20.42 & 0.642  & 27.15	& 0.0066	& 0.903	& 0.862 & 
  19.99 & 0.607 & \textcolor{red}{16.90} & \textcolor{red}{0.0019} & \textcolor{red}{0.820} & 0.881 \\
     &    \textit{Multi}-{\sc StainD}(*) (Ours)                          &  
  -- & -- & 26.02 & -- & 0.738	& 0.719 &
  -- & -- & 19.52 & -- & 0.707 & 0.783
   
    \end{tabular}}
    \vspace{-0.5em}
    \caption{\textbf{Quantitative Metrics for CK8/18 and CD3 Virtual Staining}. Models in \textcolor{blue}{Blue} represent diffusion-based approaches, with {\sc StainD} as shorthand for \name. \textbf{Bold} indicates the best performance in each model category, while \textcolor{red}{Red} highlights the second-best metric. Note that \textit{Multi}-\name is trained using only uniplex inputs but generates multiplex virtual stains, so (*) denotes cases where a corresponding IHC stain is unavailable for the H\&E image, preventing reporting of PSNR and SSIM for those outputs. Additionally, PIQ KID evaluation encountered an error, so that metric is not reported.}
    \label{tab:results_main_table}
    \vspace{-1.5em}
\end{table*}
 \textbf{Metrics:} We evaluate model performance using a suite of metrics, including two texture-based measures: PSNR and SSIM. 
 The unique paired virtual staining dataset enables reliable automated quality assessments, unlike unpaired datasets, where IHC stains are from adjacent sections or blocks requiring manual/pathologist expertise for accuracy. While metrics like PSNR and SSIM may be unreliable for unpaired datasets, they remain valid for our dataset.
We also report using four feature distribution-based metrics: FID \cite{heusel2017gans}, KID \cite{binkowski2018demystifying}, Precision, and Recall \cite{kynkaanniemi2019improved}, calculated with PIQ \cite{kastryulin2022piq} using 2048-dimensional Inception features. FID and KID measure latent space distance, assessing image quality against real images, while Precision and Recall, based on k-NN distances, estimate how well do generated samples align with the real image distribution. For a fair comparison, all metrics are reported on an image size of 128x128 for baselines and the proposed method in the main paper, results on 64x64 images and additional ablation studies are reported in the supplementary.

\subsection{Results and Discussion}

Quantitative metrics PSNR, SSIM, FID, KID, Precision, and Recall for results on CK818 and CD3 virtual staining tasks are shown in table \ref{tab:results_main_table}. 

\textbf{\name outperforms all other paired and unpaired Architectures.} Table \ref{tab:results_main_table} shows that \name consistently outperforms all other architectures for both CK8/18 and CD3 virtual staining. Among paired architectures, Bidirectional-\name ranks as the second-best model for CK8/18, while Multi-\name is the second-best for CD3 staining across most metrics. Our results demonstrate the effectiveness of the proposed architecture for multiple virtual staining tasks, including CK8/18, an epithelial cell marker, and CD3, which highlights lymphocytes that are challenging to identify by eye on H\&E.
Although Bidirectional-\name performs closely as the second- or third-best model across metrics, \name consistently outperforms it, underscoring the benefit of incorporating segmentation as an auxiliary task in multi-task training. This performance difference supports our hypothesis that segmenting cells on H\&E images and generating virtual staining conditioned on H\&E have a natural task affinity, mutually enhancing the quality of the results.
Our results also show that Multi-\name, which generates multiple stains simultaneously, achieves higher-quality outputs, with the multiplexed model ranking as the second-best for CD3 staining. The CK8/18 and CD3 markers rarely overlap—CK8/18 primarily highlights cells surrounding glandular structures, while CD3 is mostly found in the stromal regions. This separation enables Multi-\name to concentrate on distinct tissue regions, reducing common errors like incorrectly marking CD3-positive cells in glandular areas.

\begin{figure*}[!htb]
    \centering
    \includegraphics[width=1.0\linewidth]{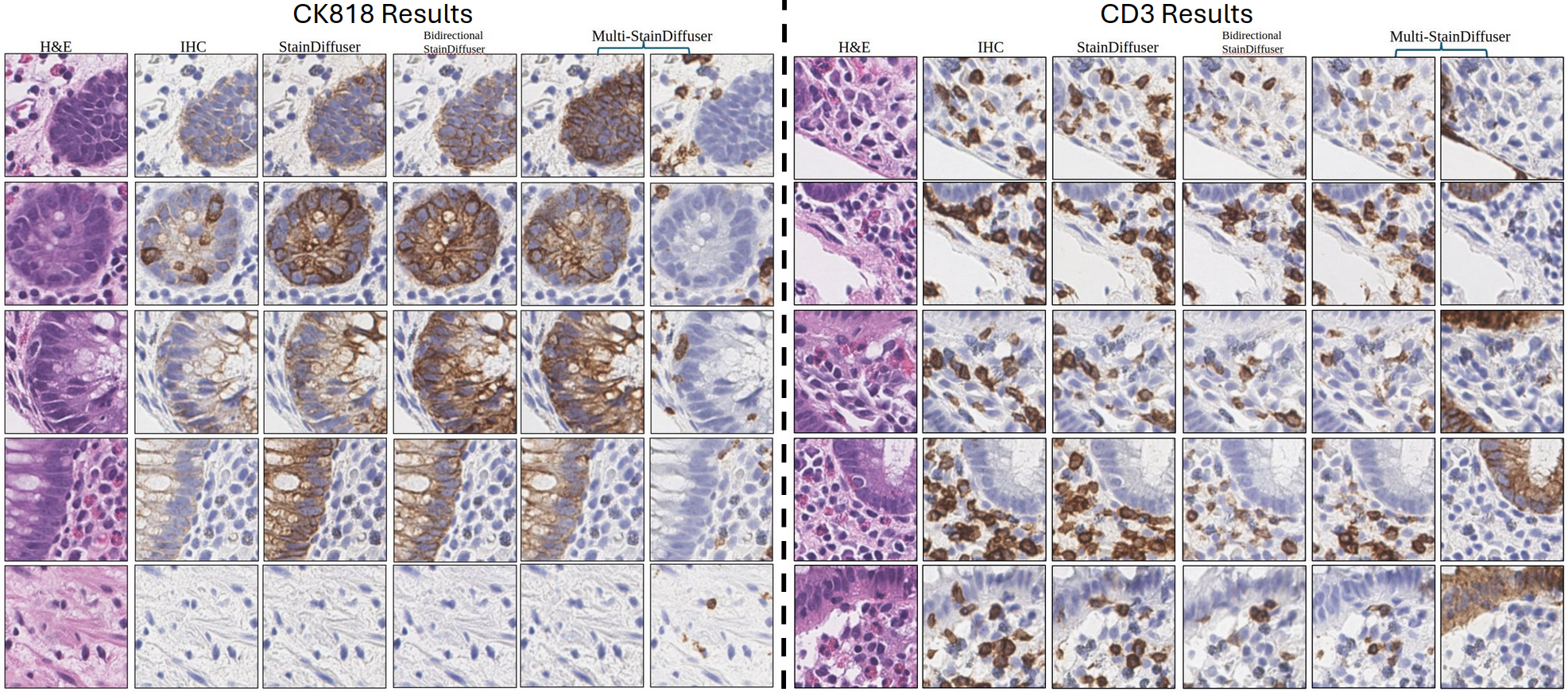}
    \vspace{-1.0em}
    \caption{\textbf{Qualitative Results.} Qualitative Results shown for CK818 and CD3 virtual staining for different models. We can observe that all the proposed models highlight the correct cells for the CK818 marker, however, \name diffuser is the best in terms of correct cell coloring for CD3 markers. }
    \label{fig:qualitativeResults}
    \vspace{-1.5em}
\end{figure*}

These results collectively demonstrate the effectiveness of multi-task learning approaches—including segmentation with conditional generation, bidirectional conditional generation, and multi-stain generation—in advancing virtual staining applications. Figure \ref{fig:qualitativeResults} presents qualitative results for different models. From the output images, it is clear that all models correctly highlight the CK8/18-positive cells, with color distribution being the primary differentiating factor. For the CD3 marker, the \name Diffuser produces the most accurate results in terms of correctly coloring the target cells, followed by Multi-\name, and then Bidirectional-\name. While some IHC stains, such as CK8/18, correspond to distinct morphological features visible in H\&E images(glandular areas), others, like CD3, do not, making virtual CD3 staining a more challenging stain. Even experienced pathologists find it challenging to differentiate between CD3+ and CD3- cells on H\&E images. Evaluating CD3 is further complicated by the need for paired H\&E and IHC images to confirm the accuracy of cell staining. The qualitative and quantitative results for CD3 underscore the difficulty of this task.

\textbf{Unpaired vs Paired Architectures.} Table \ref{tab:results_main_table} shows that paired architectures, such as PyramidPix2Pix and the proposed Stain Diffuser, achieve higher PSNR and SSIM metrics, outperforming even top unpaired models. This advantage likely stems from training on paired data, which ensures closer alignment of predicted pixel values with the original images, thus boosting texture-based metrics and explaining Stain Diffuser's strength in this area. In contrast, distribution-based metrics—such as FID, KID, precision, and recall—tend to be higher for most unpaired models. This may be because unpaired models are trained on O($N^2$) sample pairs(due to all combinations of unpaired images), enabling them to capture the dataset’s latent distribution more effectively. Notably, \name architectures are the only paired models that match or exceed unpaired models on all distribution metrics.

\textbf{Dataset Size Ablation.} Medical datasets are often in short supply, making it essential for models designed for medical applications to be robust to variations in dataset size. To assess \name's robustness, we train the model on half and a quarter of the dataset, corresponding to approximately 30,000 and 15,000 training images, respectively. Table \ref{tab:data_size_ablation} presents the results for the CD3 dataset. Our findings show that while \name performs well with half the dataset, its performance declines significantly when trained on only a quarter of the training samples. This suggests that \name can tolerate moderate reductions in dataset size but may struggle with extremely limited training data. Notably, we used the same hyperparameters for all experiments. Adjusting key parameters such as the number of diffusion steps, training epochs, or $\lambda$ for smaller datasets may lead to improved performance.

\begin{table}[!htb]
\vspace{-0.5em}
\scalebox{0.85}{
\setlength{\tabcolsep}{5pt}
    \centering
       \begin{tabular}{c|c||c|c|c|c|c}
       data & PSNR $\uparrow$ & SSIM $\uparrow$ & FID $\downarrow$& KID $\downarrow$ & Prec $\uparrow$ & Rec $\uparrow$ \\
       \hline
       Full &   20.64  & 0.633 & 15.83 & 0.0012 & 0.863 & 0.905\\
       Half & 	20.12  & 0.613 & 17.59 & 0.0023 & 0.8300 & 0.9004 \\
       Quater & 17.97 & 0.527 & 32.18 & 0.0201 & 0.7549 & 0.8339 \\
    \end{tabular}}
    \vspace{-0.5em}
    \caption{\textbf{Dataset Size Ablation.} Models trained on Half and a quarter of the training samples for CD3 dataset.}
    \label{tab:data_size_ablation}
    \vspace{-1em}
\end{table}

\textbf{Number of Diffusion Steps Ablation.} To evaluate whether the performance of virtual staining models depends on the number of diffusion steps, we trained models using step counts ranging from 500 to 4000. The results, are presented in Table \ref{tab:diffusion_steps_ablation_128} for CK818. We can observe that performance improves consistently across all metrics as the number of diffusion steps increases, with 4000 diffusion steps optimal choice for CK818.

\begin{table}[!htb]
\vspace{-0.5em}
\scalebox{0.85}{
\setlength{\tabcolsep}{5pt}
    \centering
       \begin{tabular}{c|c||c|c|c|c|c}
       Steps & PSNR $\uparrow$ & SSIM $\uparrow$ & FID $\downarrow$& KID $\downarrow$ & Prec $\uparrow$ & Rec $\uparrow$ \\
       \hline
       500  & 15.89  & 0.544 & 54.09 & 0.033 & 0.656 & 0.774 \\
       1000 & 18.11  & 0.605 & 30.18 & 0.009 & 0.832 & 0.891 \\
       2000 & 17.01  & 0.578 & 43.80 & 0.023 & 0.757 & 0.840 \\
       4000 & \bf 21.67 &	\bf 0.676 & \bf 23.46 &	\bf 0.0028 &	\bf 0.933 & \bf 0.926\\
    \end{tabular}}
    \vspace{-0.5em}
    \caption{\textbf{Diffusion Step Ablation.} Models trained with different diffusion steps on the CK818 dataset.}
    \label{tab:diffusion_steps_ablation_128}
    \vspace{-0.5em}
\end{table}

\textbf{Task Ablation.} To assess whether adding a secondary task improves model performance, we conducted ablation experiments comparing three approaches: (a) virtual staining using only one conditional diffusion process (no multi-tasking), (b) \name with a distance transform prediction as the auxiliary task, and (c) \name with segmentation as the auxiliary task. Table \ref{tab:diffusion_task_ablation_128} shows the results of these experiments for CK818 dataset. The findings indicate that incorporating an additional task, whether segmentation or distance transform, improves model performance. However, segmentation is shown to outperform compared to distance transform prediction. 

\begin{table}[!htb]
\scalebox{0.85}{
\setlength{\tabcolsep}{4.5pt}
    \centering
       \begin{tabular}{c|c|c||c|c|c|c|c}
      Seg & DT & PSNR $\uparrow$ & SSIM $\uparrow$ & FID $\downarrow$& KID $\downarrow$ & Prec $\uparrow$ & Rec $\uparrow$ \\
       \hline
      $\times$ & $\times$  & 19.89  & 0.565 & 32.99 & 0.008 & 0.787 & 0.856 \\
      $\times$ & \checkmark & 20.02 & 0.612 & 33.14 & 0.011 & 0.881 & 0.872\\
      \checkmark & $\times$  & \bf 21.67 &	\bf 0.676 & \bf 23.46 &	\bf 0.0028 &	\bf 0.933 & \bf 0.926\\
    \end{tabular}}
    \vspace{-0.5em}
    \caption{\textbf{Task Ablation.} Models trained with different auxiliary tasks: no multi-tasking, segmentation, or distance transform.}
    \label{tab:diffusion_task_ablation_128}
    \vspace{-1.0em}
\end{table}

\textbf{Image Resolution Ablation.} We trained models on images of varying resolutions to determine whether certain stains can be effectively generated at lower resolutions without sacrificing output quality.  The results of these experiments are shown in Supplementary Table \ref{tab:results_resolution_differences}. For CK8/18 virtual staining, lower-resolution images improved all performance metrics. However, for CD3, using lower resolution reduced model performance. These findings suggest that some stains can be generated effectively at lower resolutions, saving computational resources without compromising accuracy.  Our \name model on 128x128 images takes four times longer per epoch than higher resolutions, so training CK818 at lower resolutions saves GPU resources without sacrificing performance.

\textbf{Loss Weight Ablation($\lambda$).} To assess the impact of the $\lambda$ hyperparameter in the \name loss function, we trained models with various $\lambda$ values: 0.1, 0.3, 1, 3, and 10 (64x64 image size to save compute). Results shown in Table \ref{tab:lambda_ablation} for CK818 and table \ref{tab:lambda_ablation_cd3} for CD3(in Supplementary), show that $\lambda = 1 $ or $3$ yields the best-performing model, suggesting that equal weighting of tasks is optimal for CD3 but CK8/18 virtual staining performs better with unequal weights between segmentation and generation task. 

\textbf{Limitations.} While our model is flexible in generating multiple stains, scalable, and capable of producing high-quality virtual stains as demonstrated in results section, it has certain limitations. As shown in Table \ref{tab:results_main_table} and \ref{tab:task_ablation}, \name’s performance is influenced by the auxiliary task used, with segmentation yielding the best results. This dependency poses challenges for stains where segmentation via thresholding or morphological operations \cite{kataria2023automating} isn’t feasible. Additionally, \name is tailored for paired datasets (H\&E and IHC stains on the same tissue), which may not be universally available, making data acquisition protocols a potential constraint. Our study also does not assess \name’s sensitivity to registration accuracy between H\&E and IHC images, leaving this as a direction for future research. Although diffusion models provide high sampling diversity and quality, they are slower during inference, potentially limiting use in resource-constrained settings—an issue that could be addressed by exploring fast-sampling diffusion methods \cite{zheng2023fast, xiao2021tackling}. Lastly, because the models were trained on data from a single site, they may experience performance degradation on data from other sites due to input distribution shifts.
\section{Conclusion and Future Work}
We propose a novel multi-task diffusion model architecture that demonstrates effectiveness and robustness in virtual staining applications. The model uses multi-task diffusion processes to learn conditional image-to-image (I2I) tasks: (a) accurately staining cells for H\&E conditioned immunohistochemistry generation (IHC), and (b) simultaneously segmenting the same cells on H\&E input image, creating an implicit affinity between the tasks. Leveraging the flexibility of our architecture, we extend it to multiplex staining, enabling the generation of multiple stains on the same H\&E image using a single model trained on uni-plex data. We conduct extensive comparisons with both paired and unpaired I2I models, showing that our approach outperforms numerous methods in the literature. These comparisons establish new benchmarks for virtual staining, offering valuable insights for future research in the field. In future work, we aim to extend Multi-\name to handle multiple stains, creating a foundational model for virtual staining, and explore alternative attention mechanisms \cite{dao2022flashattention,shah2024flashattention,wang2020linformer} to improve training efficiency for resource and data-constrained environments.
{
    \small
    \bibliographystyle{ieeenat_fullname}
    \bibliography{main}
}
\clearpage
\setcounter{page}{1}
\maketitlesupplementary
\section{Diffusion model Background}\label{apdx:diffusion_model_background}
Here, we explain background related to diffusion models and conditional diffusion models. 
The denoising diffusion probabilistic model (DDPM) defines the diffusion process as a two-stage process, i.e., forward and reverse. 
The \textit{forward process} progressively corrupts the image, $\textbf{I}_0 \in \mathbb{R}^{H \times W}$, by adding Gaussian noise over $T$ iterations in a Markovian manner. Gaussian noise is added at each step according to a variance schedule defined by $\beta_1, \cdots, \beta_T$.
The noisy image at time step $t$, denoted by $\textbf{I}_{t}$, can be sampled from the following conditional distribution: 
\vspace{-0.5em}
\begin{equation}\label{eqn:conditional}
    q(\textbf{I}_{t}|\textbf{I}_{t-1})= \mathcal{N}(\textbf{I}_t;\sqrt{1-\beta_t}\textbf{I}_{t-1},\beta_t {\mathbb{I}})
\end{equation}
where $\mathbb{I}$ is the identity matrix. The variance schedule $\{\beta_t\}_t=1^T$ is chosen such that after $T$ steps, $\textbf{I}_T$ becomes indistinguishable from pure Gaussian noise. 
Marginalizing Eq. \ref{eqn:conditional} over $t$ 
gives the corruption equations at an arbitrary time step $t$:
\vspace{-0.5em}
\begin{equation}
    q(\textbf{I}_{t}|\textbf{I}_0)= \mathcal{N}(\textbf{I}_t;\sqrt{\Bar{\alpha}}\textbf{I}_0,(1-\Bar{\alpha})\mathbb{I})
\end{equation}
where $\Bar{\alpha}_t = \prod_{s=1}^t {\alpha}_s$, and $\alpha_t = 1-\beta_t$. This equation provides a closed-form solution for sampling in the forward process at an arbitrary time step $t$. This Gaussian re-parameterization also gives a closed-form formulation of the posterior distribution of $\textbf{I}_{t-1}$ given ($\textbf{I}_0,\textbf{I}_t$)
\vspace{-0.5em}
\begin{align}
    q(\textbf{I}_{t-1}|\textbf{I}_0,\textbf{I}_t)&= \mathcal{N}(\textbf{I}_{t-1};\boldsymbol{\mu}_t(\mathbf{I}_t, \mathbf{I}_0),{\sigma_t}\mathbb{I}), \text{where}   \\
   \boldsymbol{\mu}_t(\mathbf{I}_t, \mathbf{I}_0)&=\frac{\sqrt{\Bar{\alpha}_{t-1}}(\beta_t)}{1-\Bar{\alpha}_{t}}\mathbf{I}_{0}+\frac{\sqrt{\Bar{\alpha}_{t}}(1-\Bar{\alpha}_{t-1})}{(1-\Bar{\alpha}_t)}\mathbf{I}_{t} \\
   {\sigma}_t&=\frac{(1-\Bar{\alpha}_{t-1})\beta_t}{(1-\Bar{\alpha}_{t})}
\end{align}
Equations 3,4 and 5 are used during inference to generate images from Gaussian noise. The \textit{reverse process} learns to denoise the noisy image using a deep neural network. 
A neural network parameterized by $\mathbf{\theta}$, denoted as $f_{\theta}( \textbf{I}_t(\mathbf{I}_0,\Bar{\alpha}_t),t)$, takes the noisy image $\mathbf{I}_t$ and the current random noise level $\Bar{\alpha}_t$ as inputs to estimate the noise vector $\boldsymbol{\epsilon}$ used to corrupt the original image $\mathbf{I}_0$. The loss function is defined as the mean squared error (MSE) between the estimated noise and the original noise added, as described in Equation 2.
\vspace{-0.5em}
\begin{align}
    \mathcal{L}(\theta) &= \mathbb{E}_{t, \mathbf{I}_0, \boldsymbol{\epsilon}} \left[||\boldsymbol{\epsilon}-f_{\theta}( \textbf{I}_t(\mathbf{I}_0,\Bar{\alpha}_t),t)||^2 \right] 
\end{align}
where
\vspace{-0.5em}
\begin{align}
    \textbf{I}_t(\mathbf{I}_0,\Bar{\alpha}_t) &= \sqrt{\Bar{\alpha}_t} \mathbf{I}_0 + \sqrt{1-\Bar{\alpha}_t}\boldsymbol{\epsilon}, \text{~for~} \boldsymbol{\epsilon} \sim \mathcal{N}(\mathbf{0}, \mathbb{I})
\end{align}
In conditional diffusion models, the network, parameterized by $\theta$, is also conditioned on the input image $\mathbf{I}_x$ or a latent space representation from a pre-trained model, resulting in an estimation function defined as
$f_{\theta}( \mathbf{I}_x, \textbf{I}_t(\mathbf{I}_0,\Bar{\alpha}_t),t)$. The loss function for conditional diffusion models is given by:
\vspace{-0.5em}
\begin{align}
    \mathcal{L}_{\text{cond}}(\theta) &= \mathbb{E}_{t, (\mathbf{I}_x, \mathbf{I}_0), \boldsymbol{\epsilon}} \left[||\boldsymbol{\epsilon}-f_{\theta}( \mathbf{I}_x, \textbf{I}_t(\mathbf{I}_0,\Bar{\alpha}_t),t)||^2 \right] 
\end{align}

\section{Additional Data Acquisition and Implementation Details} \label{apdx:data_aquistion}
\textbf{Data Acquisition Details.} 
The dataset includes H\&E whole slide images from surveillance colonoscopies of patients with active ulcerative colitis, which contained 92 tissue pieces. The slides were stained with H\&E using an automated clinical staining process and scanned on the Aperio AT2 slide scanner at a pixel resolution of 0.23 µm at 40x magnification. After scanning, the coverslips were removed, and the slides were restained using the Leica Bond 3 autostainer with antibodies against CD3 (cluster of differentiation 3) or CK8/18 (cytokeratin-8/18). The heat retrieval step before antibody incubation decolorized the H\&E slides, eliminating the need for manual removal. After IHC staining, the slides were rescanned on the Aperio AT2 at 40x, and the digital IHC images were paired with the corresponding H\&E images. Due to the use of different scanners, registration is necessary to align the tissue patches. We employed the multi-resolution framework of ANTS for tissue piece registration. The tissue alignments were manually verified and adjusted until the desired alignment accuracy was achieved.

\textbf{Automated Coarse Segmentation.} We used similar methodology to one explained in \citet{kataria2023automating} for automated annotation for CK8/18 stain. However, for CD3 stain we adapted some components, such as thresholding the DAB channel was done using a different thresholding operations
\begin{align}
    threshold = \mu +2* \sigma
\end{align}
where $\mu$ and $\sigma$, the values of the DAB for the WSI. Through manual observation we noted that CD3 stain segmentation after thresholding was not consistent due to stain variations. To make the segmentation more consistent,   we trained a segmentation model, with noisy labels and then used the predicted segmentation from the model as our coarse segmentation.

\textbf{Additional Implementation Details.} All baselines were trained using default parameters from their original GitHub repositories, with minimal changes for compatibility.  All models were trained on NVIDIA A100 40GB GPU. Additional relevant details for specific baselines:-
\begin{itemize}
    \item For CycleDiffusion \cite{wu2022unifying} and EGSDE \cite{zhao2022egsde}, intermediate diffusion models were trained for 160,000 iterations with guided diffusion on TitanX 12 GB GPU. 
    \item For LBBDM, we used the default VQ-VAEs provided with the original implementation was used for training the models. Domain specific VQ-VAEs might perform better, but that is left for future work.
    \item For SANTA, a batch size of 4 resulted in the model throwing an error. So these models were trained with the default batch size in the original implementation.
    \item Distance Transform implementation was borrowed from skimage python library.
\end{itemize}

\section{Ablation Results on 64x64 images}

To conserve computational resources, we conducted extensive ablation studies at 64x64 resolution to assess the impact of various hyperparameters and design choices; the results and analysis are presented here.

\textbf{Image Resolution Ablation.} We trained models on images at varying resolutions to assess whether certain stains could be effectively generated at lower resolutions without compromising quality. The results, detailed in Supplementary Table \ref{tab:results_resolution_differences}, reveal that for CK8/18 virtual staining, lower-resolution images improved all performance metrics. In contrast, CD3 performance declined with reduced resolution. These findings indicate that some stains can be reliably generated at lower resolutions, offering significant computational savings. For instance, training \name on 128x128 images takes four times longer per epoch than lower-resolution models. Thus, training CK8/18 at reduced resolutions can conserve GPU resources while maintaining performance. 
For future extension of this idea, we want to explore whether we get similar high performance by training super-resolution models to convert low resolution images to higher resolution and still get superior performance. Additionally we also want to explore if both \name and super-resolution models can be trained together, that way an end-end training will make other smaller resolution like 32x32 feasible and reduce the amount of compute needed to train \name architectures. 

\begin{table*}[!htb]
\setlength{\tabcolsep}{1pt}
    \centering
    \begin{tabular}{c|c||c|c||c|c|c|c||c|c||c|c|c|c}
    & & \multicolumn{6}{c||}{\bf CK818} &  \multicolumn{6}{c}{\bf CD3}
    \\
    \hline
    & & \multicolumn{2}{c||}{\bf Texture Metrics } &  \multicolumn{4}{c||}{\bf Distribution Metrics} 
   & \multicolumn{2}{c||}{\bf Texture Metrics } &  \multicolumn{4}{c}{\bf Distribution Metrics}  \\
    \hline
  ImageSize  & Method & PSNR $\uparrow$ & SSIM $\uparrow$ & FID $\downarrow$& KID $\downarrow$ & Prec $\uparrow$ & Rec $\uparrow$  & PSNR $\uparrow$ & SSIM $\uparrow$ & FID $\downarrow$& KID $\downarrow$ & Prec $\uparrow$ & Rec $\uparrow$ \\
   \hline
  64x64 &   \name                                     &  
     \bf 23.08 &	\bf 0.782 & \bf 22.09 &	\bf 0.0018 &	\bf 0.927 & \bf 0.931 &
   \bf   20.70    &  \bf 0.722 & \bf 18.03 & \bf 0.0019 & \bf 0.843 & \bf 0.904 \\
  64x64 &    \textit{BiDirec}-{\sc StainD}                    &  
   \textcolor{red}{22.96}	& \textcolor{red}{0.782}	& \textcolor{red}{22.78} &	\textcolor{red}{0.0024}	& \textcolor{red}{0.918}	& \textcolor{red}{0.9299} &
   21.36 & 0.722 & 18.03 & 0.0019 & 0.852 & \textcolor{red}{0.905} \\
 64x64  &    \textit{Multi}-{\sc StainD}                          &  
  22.79 & 0.769  & 23.32	& 0.0042	& 0.905	& 0.9159 & 
  21.12 & 0.700 & \textcolor{red}{19.19} & \textcolor{red}{0.0034} & \textcolor{red}{0.843} & 0.885 \\
      &    \textit{Multi}-{\sc StainD}(*)                          &  
  -- & -- & 26.32 & -- & 0.735	& 0.800 &
  -- & -- & 20.86 & -- & 0.710 & 0.792 \\
  \hline
  128x128 &   \name                                     &  
     \bf 21.67 &	\bf 0.676 & \bf 23.46 &	\bf 0.0028 &	\bf 0.933 & \bf 0.926 &
   \bf   20.64    &  \bf 0.633 & \bf 15.83 & \bf 0.0012 & \bf 0.863 & \bf 0.905 \\
  128x128 &    \textit{BiDirec}-{\sc StainD}                    &  
   \textcolor{red}{20.47}	& \textcolor{red}{0.650}	& \textcolor{red}{25.02} &	\textcolor{red}{0.0047}	& \textcolor{red}{0.914}	& \textcolor{red}{0.880} &
   20.08 & 0.600 & 19.22 & 0.0034 & 0.779 & \textcolor{red}{0.891} \\
 128x128  &    \textit{Multi}-{\sc StainD}                          &  
  20.42 & 0.642  & 27.15	& 0.0066	& 0.903	& 0.862 & 
  19.99 & 0.607 & \textcolor{red}{16.90} & \textcolor{red}{0.0019} & \textcolor{red}{0.820} & 0.881 \\
     &    \textit{Multi}-{\sc StainD}(*)                          &  
  -- & -- & 26.02 & -- & 0.738	& 0.719 &
  -- & -- & 19.52 & -- & 0.707 & 0.783

 \end{tabular}
    \caption{\textbf{Quantitative Metrics for CK8/18 and CD3 Virtual Staining Comparing models trained at different resolutions}. {\sc StainD} as shorthand for \name. \textbf{Bold} indicates the best performance in each model category, while \textcolor{red}{Red} highlights the second-best metric. Note that \textit{Multi}-\name is trained using only uniplex inputs but generates multiplex virtual stains, so (*) denotes cases where a corresponding IHC stain is unavailable for the H\&E image, preventing reporting of PSNR and SSIM for those outputs. Additionally, PIQ KID evaluation encountered an error, so that metric is not reported.}
    \label{tab:results_resolution_differences}
\end{table*}

\textbf{Task Ablation at 64x64 image size.} Similar to task ablation in main paper Table \ref{tab:diffusion_task_ablation_128}, we also ran task ablation for both dataset on 64x64 images. Table \ref{tab:task_ablation}, shows the results of these ablation experiments. The findings indicate that incorporating an additional task, whether segmentation or distance transform, improves model performance even on small resolution generation task. Specifically, the segmentation task yields better results for virtual staining of CK818, while the distance transform task achieves superior performance for CD3. 
\begin{table}[!htb]
\setlength{\tabcolsep}{2pt}
\scalebox{1.0}{
    \centering
    \begin{tabular}{p{0.2cm}|c|c||c|c|c|c|c|c}
      &  seg & DT & PSNR & SSIM & FID & KID & Prec & Rec \\
        \hline
        \multirow{3}{=}{\begin{sideways}\textbf{CK818}\end{sideways}} 
        & $\times$ & $\times$ & 21.16 & 0.748 &  26.8 & 0.0041 & 0.82 & 0.784 \\
         & $\times$ & \checkmark    &  22.01 & 0.762 & 25.63 & 0.0043 & 0.917 & 0.911\\
       & \checkmark &$\times$   &  \bf 22.83 & \bf 0.775 & \bf 22.86 & \bf 0.0027 & \bf 0.918 & \bf 0.926\\
        \hline
        \multirow{3}{=}{\begin{sideways}\textbf{CD3}\end{sideways}}  
        & $\times$ & $\times$ &  19.64 & 0.614 & 38.64 & 0.0264 &  0.702 & 0.679\\
        & $\times$ & \checkmark        &  \bf 20.65 & \bf 0.681 & \bf 18.94 & \bf 0.0026 & \bf 0.829 & 0.8838\\
        & \checkmark &$\times$      &  20.64 & 0.659 & 23.02 & 0.0072 & 0.765 & \bf 0.903\\ 
    \end{tabular} }
    \caption{\textbf{Auxiliary Task Ablation.} We performed experiments to assess the effect of various auxiliary tasks on the performance of \name in virtual staining. The results indicate that incorporating segmentation as an auxiliary task yields the highest model performance.}
    \label{tab:task_ablation}
\end{table}

\textbf{Diffusion Steps Ablation at 64x64 image size.} Similar to our findings in Table \ref{tab:diffusion_steps_ablation_128}, we conducted an ablation study on diffusion steps at lower resolution 64x64 (see Table \ref{tab:diffusion_steps_ablation}). In general, increasing the number of diffusion steps improves quantitative performance, although the trends at lower resolution are more nuanced. For CD3 virtual staining models, fewer diffusion steps yield reasonable SSIM, precision, and recall, while models with over 2000 steps achieve the best FID and PSNR scores. These results suggest that, for some virtual staining models, using fewer diffusion steps can still produce competitive performance.

\textbf{Loss Weight Ablation Results for CD3.} To evaluate the impact of the $\lambda$ hyperparameter in the \name loss function, we trained models on the CK818 and CD3 dataset using a range of $\lambda$ values: 0.1, 0.3, 1, 3, and 10. Results are shown in Table \ref{tab:lambda_ablation_cd3} and \ref{tab:lambda_ablation}. When comparing with results in Table \ref{tab:lambda_ablation}, it becomes evident that increasing the weight of the segmentation branch enhances the generation branch's accuracy for CD3 staining. This finding suggests that for certain stains, such as CD3, where staining features are less distinct in the H\&E image, emphasizing the segmentation branch contributes to improved staining accuracy.

\begin{table}[!htb]
\scalebox{1.0}{
    \centering
    \begin{tabular}{c||c|c|c|c|c|c}
        $\lambda$ & PSNR & SSIM & FID & KID & Prec & Rec \\
        \hline
        0.1 &  22.19 & 0.749 & 25.20 & 0.0043 & 0.872 & 0.918\\
        0.3 &  22.44 & 0.760 & 25.31 & 0.0044 & 0.886 & 0.918\\
        1   &  \bf 22.83 & \bf 0.775  &  \bf 22.86 &  \bf 0.0027 & \bf 0.918 & \bf 0.926\\
        3   &  21.44 & 0.746  & 28.23 & 0.0065 & 0.871 & 0.895\\
        10  &  20.32 & 0.713  & 35.59 & 0.0138 & 0.831 & 0.857 \\
    \end{tabular} }
    \vspace{-0.5em}
    \caption{\textbf{Loss Weighting Ablation}. Results demonstrate the impact of varying the $\lambda$ hyperparameter in the loss function.}
    \label{tab:lambda_ablation}
    \vspace{-1.em}
\end{table}

\begin{table}[!htb]
\scalebox{0.9}{
    \centering
    \begin{tabular}{c||c|c|c|c|c|c}
        $\lambda$ & PSNR $\uparrow$& SSIM $\uparrow$ & FID $\downarrow$& KID $\downarrow$ & Prec $\uparrow$ & Rec $\uparrow$\\
        \hline
        0.1 &  20.89 & 0.673  & 21.94  & 0.0060  & 0.758 & 0.911\\
        0.3 &  17.51 & 0.578  & 46.90  & 0.0278  & \bf 0.802 & 0.781\\
        1   &  20.64 & 0.659  & 23.02  & 0.0072  & 0.765 & 0.903\\ 
        3   &  \bf 21.37 & \bf 0.708  & \bf 19.95  & \bf 0.0036 & 0.795 & \bf 0.906\\
        10  &  21.04 & 0.700  & 25.50  & 0.0078 & 0.741 & 0.904\\
    \end{tabular} }
    \caption{\textbf{Loss Weighting Ablation For CK818}. Results demonstrate the impact of varying the $\lambda$ hyperparameter in the loss function.}
    \label{tab:lambda_ablation_cd3}
\end{table}

\begin{table}[!htb]
    \centering
    \setlength{\tabcolsep}{4pt}
    \scalebox{1.0}{
    \begin{tabular}{p{0.15cm}|c||c|c|c|c|c|c}
       & steps & PSNR & SSIM & FID & KID & Prec & Rec \\
        \hline
        \multirow{5}{=}{\begin{sideways}\textbf{CK818}\end{sideways}}  
       & 250   &  22.97 & \bf 0.782    & 27.25 & 0.0062 & 0.906 & 0.91\\
       & 500   &  22.70 & 0.772        & 25.02 & 0.0044 & 0.913 & 0.918\\
       & 1000  &  22.83 & 0.775        & 22.86 & 0.0027 & 0.918 & 0.926\\
       & 2000  &  21.68 & 0.747        & 27.63 & 0.0064 & 0.9   & 0.897\\
       & 4000  &  \bf 23.08 & \bf 0.782& \bf 22.09 & \bf 0.0018 & \bf 0.927 & \bf 0.931\\
        \hline
        \multirow{5}{=}{\begin{sideways}\textbf{CD3}\end{sideways}}  
       & 250   &  21.33 & \bf 0.714 & 18.77 & 0.0024 & \bf 0.844 & \bf 0.907\\
       & 500   &  20.57 & 0.678     & 22.05 & 0.0056 & 0.841 & 0.879\\
       & 1000  &  20.64 & 0.659     & 23.02 & 0.0072 & 0.765 & 0.903\\ 
       & 2000  &  20.70 & 0.680     & \bf 18.33 & \bf 0.0017 & 0.843 & 0.904\\
       & 4000  &  \bf 21.38 & 0.708 & 22.43 & 0.0056 & 0.821 & 0.897\\
    \end{tabular} }
    \vspace{-0.5em}
    \caption{\textbf{Diffusion Steps Ablation on 64x64 image size.} Results when changing the number of diffusion steps in  \name.}
    \label{tab:diffusion_steps_ablation}
    \vspace{-1.em}
\end{table}

\section{Additional Qualitative Results}
\begin{figure*}[!htb]
    \centering
    \includegraphics[width=0.7\linewidth]{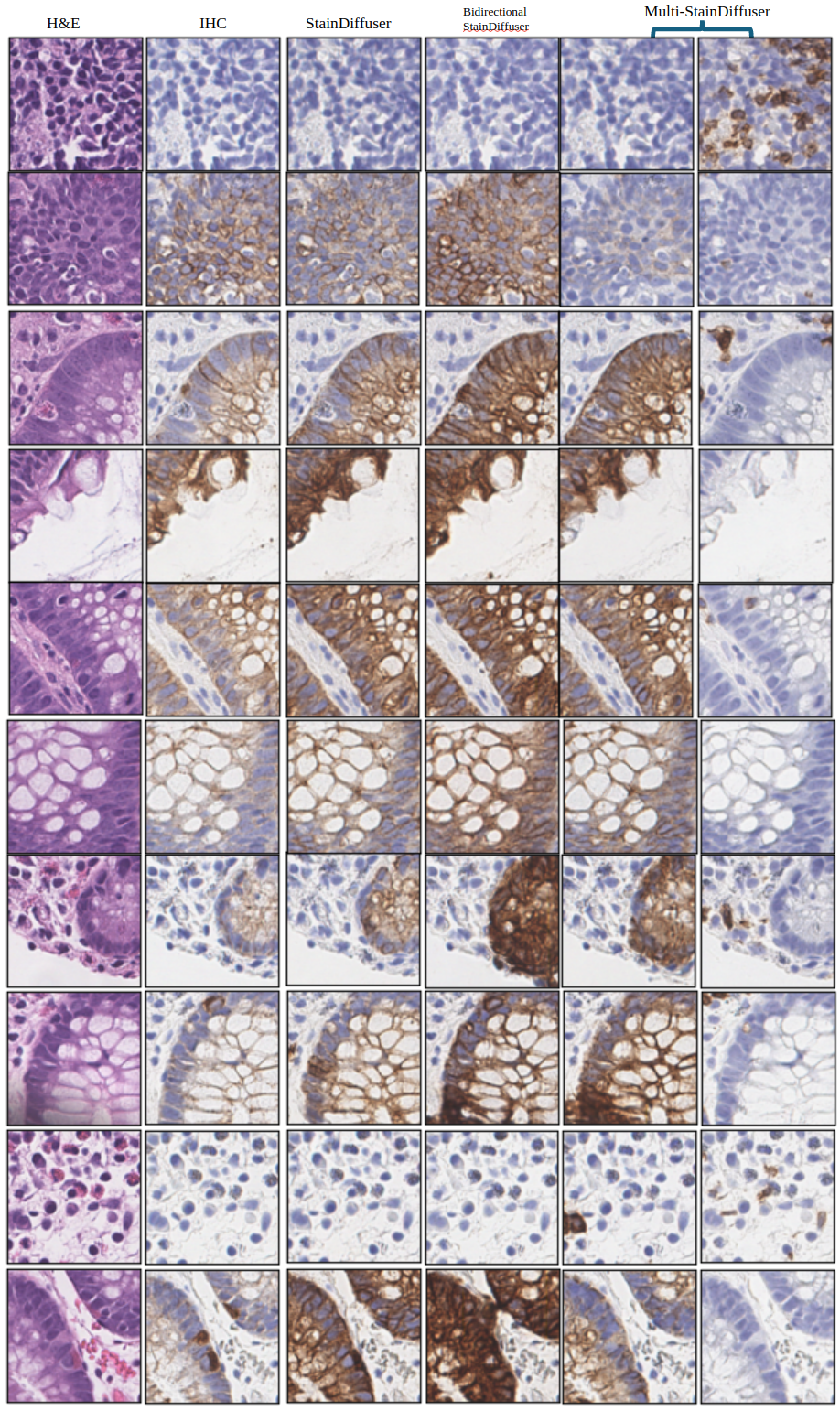}
    \caption{\textbf{Additional CK818 Qualitative Results Comparing Proposed Methods}. The results show that all proposed methods perform reasonably well in achieving accurate staining. However, \name stands out with the highest precision in exact color matching.}
    \label{fig:ck818_qualitative_results_supp}
\end{figure*}

\begin{figure*}[!htb]
    \centering
    \includegraphics[width=0.7\linewidth]{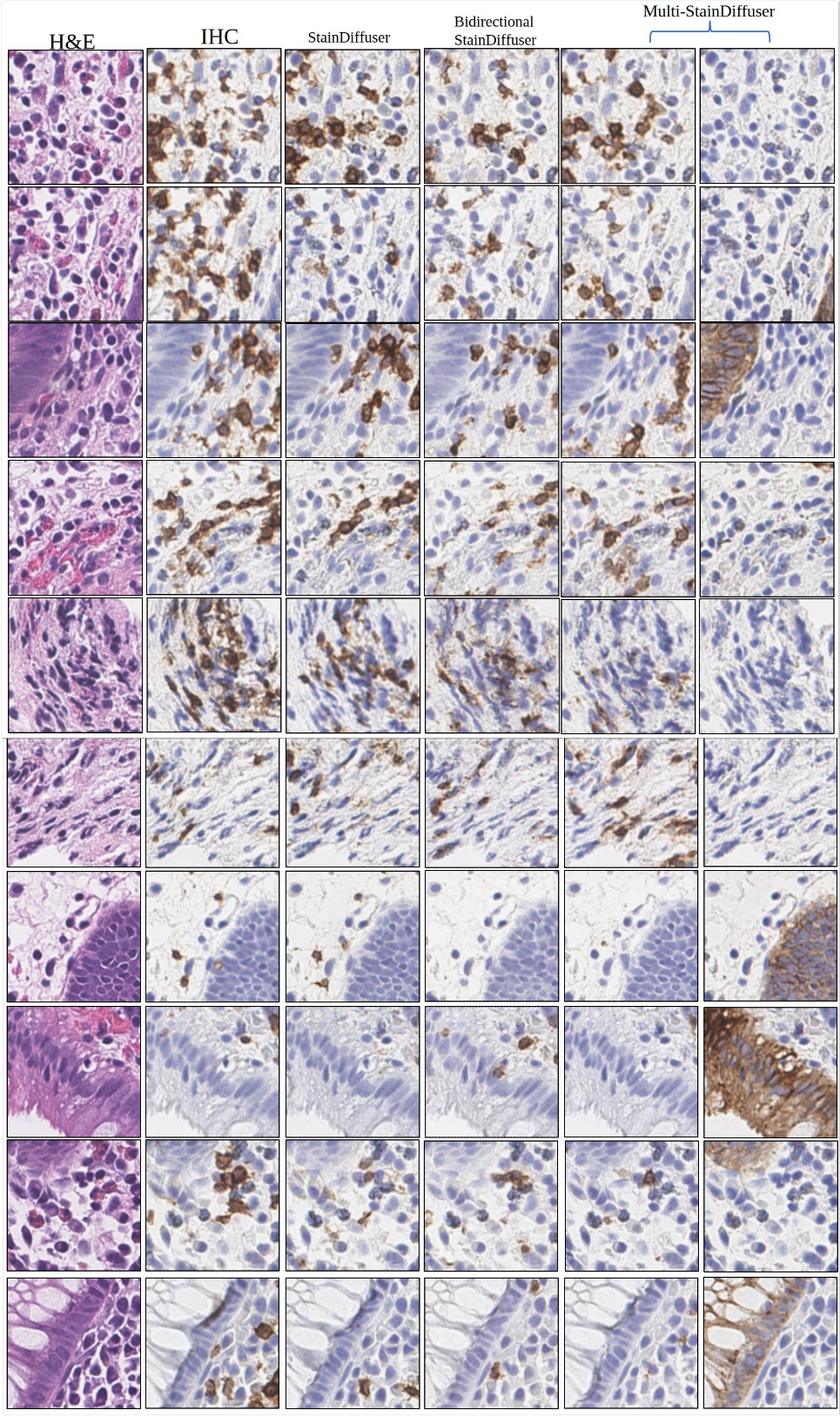}
    \caption{\textbf{Additional CD3 Qualitative Results Comparing Proposed Methods}. \name stands out by achieving the highest precision in accurately coloring the greatest number of correct cells.}
    \label{fig:cd3_qualitative_results_supp}
\end{figure*}

Additional qualitative results comparing \name, Bi-directional-\name, and Multi-\name are shown in Figure \ref{fig:ck818_qualitative_results_supp} and \ref{fig:cd3_qualitative_results_supp}, for CK818 and CD3 virtual staining respectively. From figure \ref{fig:ck818_qualitative_results_supp}, it is evident that all models perform reasonably well in staining the correct cells for CK818. However, Multi-\name exhibits two notable issues: in the 9th row, it incorrectly stains an additional cell, and in Sample 2, the stain appears less intense, resulting in a lower-quality output compared to the others.  Bi-directional-\name results in a darker stain compared to the original stain, but highlights the correct cells in all cases. 

From figure \ref{fig:cd3_qualitative_results_supp}, \name demonstrates the highest accuracy in staining the correct cells across all cases. Notably, in Sample 8 (Row 8), \name successfully highlights even a small number of CD3+ cells, a feat that other methods fail to achieve. These results further validate the efficacy of leveraging task affinity between segmentation and virtual staining, a feature absent in Bi-Direction and multi-stain approaches.

\end{document}